\documentclass[twocolumn,preprintnumbers,amsmath,amssymb,floatfix]{revtex4}
\usepackage{graphicx, subfigure}% Include figure files
\usepackage{dcolumn}% Align table columns on decimal point
\usepackage{bm}% bold math
\usepackage{epsf}
\usepackage{amsmath}
\usepackage{amsfonts}
\newcommand{\DDir}{\relax{D\kern-.7em{/}}}

\newcommand{\inv}[1]{\frac{1}{#1}}

\newcommand{\ra}{\rightarrow}

\newcommand{\be}{\begin{equation}}
\newcommand{\ee}{\end{equation}}
\newcommand{\bea}{\begin{equation*}}
\newcommand{\eea}{\end{equation*}}

\newcommand{\abs}[1]{\left\vert#1\right\vert}

\newcommand{\nin}{\relax{\in\kern-.8em{/}}}

\newcommand{\om}{\omega}

\newcommand{\cm}{\mbox{ cm}}

\newcommand{\yr}{\mbox{ yr}}

\newcommand{\Gyr}{\mbox{ Gyr}}

\newcommand{\km}{\mbox{ km}}

\newcommand{\Mpc}{\mbox{ Mpc}}

\newcommand{\sref}{\S~\ref}

\newcommand{\aj}{AJ}
\renewcommand{\apj}{ApJ}
\newcommand{\apjl}{ApJ}

\newcommand{\aap}{A\&A}

\newcommand{\Min}{m_1+m_2}
\newcommand{\Minp}{(m_1+m_2)}
\newcommand{\ADJ}{2.55}%[15\pi]^{1/3}/sqrt(2)
\newcommand{\apjs}{Astrophysical Journal Supp.}
\newcommand{\mnras}{MNRAS}
\begin{document}
\title{The rate of White Dwarf-White Dwarf head-on collisions may be as high as the type Ia supernova rate}

\author{Boaz Katz$^{1 *}$, Subo Dong$^{1}$}
\affiliation{$^1$Institute for Advanced Study, Princeton, NJ 08540, USA}
\begin{abstract}
We show that a White Dwarf-White Dwarf (WD-WD) binary with semi-major axis $a\sim1-300$AU, which is orbited by a stellar mass outer perturber with a moderate pericenter $r_{p,\rm out}/a\sim 3-10$,  has a few percent chance of experiencing a head-on collision within $\sim 5\Gyr$. Such a perturber is sufficiently distant to allow the triple system to remain intact for millions of orbits yet close enough to efficiently exchange angular momentum with the WD-WD binary.  In $\sim 5\%$ of initial orientations, the inner orbit stochastically scans the phase space in the neighborhood of zero angular momentum. In these systems, the binary experiences increasingly closer pericenter approaches $r_p\sim a/2N$ with the increasing number ($N$) of orbits elapsed. Within $N\sim 10^5(a/30\rm AU)$ orbits, a collision is likely to occur. This is shown by performing $\sim$ten thousand 3-body integrations and is explained by simple analytic arguments. The collisions are conservatively restricted to 'clean' collisions in which all passages prior to the collision are greater than $4R_{\rm WD}=4\times 10^9\cm$. In particular, in the last single orbit, the pericenter ``jumps'' from $r_p>4R_{\rm WD}$ to a collision value of $r_p<2R_{\rm WD}$. The effects of tidal deformations and General Relativistic (GR) corrections are negligible in these scenarios.  The WDs approach each other with a high velocity $>3000$ km/s and the collision is likely to detonate the WDs leading to a type Ia SN. If a significant fraction of WDs reside in such triples, the rate of such collisions is as high as the SN Ia rate, and it is possible that some or all type Ia SNe occur in this way. Such SNe have a unique gravitational wave signature, which will allow a decisive identification in the future.  
\end{abstract}
\maketitle
\section{Introduction}\label{sec:Introduction}
The merger of two white dwarfs (WDs) due to gravitational energy loss in close WD binaries (the double degenerate scenario) is one of the leading mechanisms for producing type Ia supernovae (see e.g.  \cite{Howell11} for a review). For the WD-WD merger scenario to work, two challenging conditions must be met:
1. The merger rate is sufficient.
2. A successful detonation occurs during the merger.
Neither condition has been established so far for this scenario. 

We study an alternative double degenerate scenario wherein the two WDs collide head-on, which has not yet been considered as a primary channel of type Ia supernova (SN) production. In fact, it is very likely that such a collision would detonate the WDs for typical, sub-Chandrasekhar $0.6M_{\odot}-0.6M_{\odot}$ collisions \cite{Rosswog09,Raskin09,Raskin10,Hawley12} (see however \cite{Loren10}). Indeed, while approaching a collision, the WDs reach a high relative velocity $v\sim3600 [(m_1+m_2)/M_{\odot}]^{0.5}[(R_1+R_2)/2\times10^9\cm]^{-0.5} \rm km/s$ (where $m_1,R_1$ and $m_2,R_2$ are the mass and radius of the WDs) and the resulting shock waves are likely to trigger a thermonuclear explosion producing a type Ia SN. 

The main objection to the head-on collision scenario is the common perception that such collisions are extremely rare. The collisions are believed to predominantly occur in dense stellar environments, such as cores of globular clusters, with a rate that is orders of magnitude smaller than the SN Ia rate of $3\times 10^{-5}\Mpc^{-3}\yr^{-1}$ (e.g. \cite{Rosswog09,Raskin09}). We show that this perception is wrong: The rate of WD-WD collisions may be as high as the SN Ia rate with the collisions occurring in field triple star systems with inner WD-WD binary orbital separations of $a \sim1-300\,\rm AU$ (typical for field binaries) due to the excitation of the inner orbital eccentricities to values extremely close to unity $1- e \sim 10^{-6}$.

Orbital eccentricities can be driven to high values due to the perturbation from highly inclined tertiaries (perturbers) in hierarchical triple systems\cite{Lidov62,Kozai62}. In this so-called  Kozai-Lidov mechanism, the inner orbital eccentricity oscillates periodically and slowly, on a secular time scale much greater than the inner and outer orbital periods. The Kozai-Lidov mechanism allows 
distant astrophysical objects to be driven into close 
pericenter approaches.  In particular, such close approaches have 
been invoked to allow energy dissipative mechanisms (e.g. tidal dissipation or gravitational radiation) to operate efficiently and 
produce tight stellar binaries (e.g., \cite{eggleton, fabrycky, dong}), 
black hole mergers (e.g., \cite{blaes}), hot Jupiters (e.g., \cite{wumurray, fabrycky, socrates}), etc. In the context of SNe Ia production,  it has been recently suggested that a similar process may 
enhance the WD-WD merger rate due to gravitational radiation  \cite{Thompson11,Shappee12}.
   
However, direct collisions of WDs due to Kozai-Lidov mechanism 
have not been considered as a serious possibility. There are three challenges leading to the perception that such collisions in triple systems are extremely rare or impossible: a. the daunting distance scale ratio involved  $a/R_{\rm WD}\sim 10^6(a/30\rm AU)$ which corresponds to an extreme required eccentricity $1-e>10^{-6}$, b. a common concern that tidal and General Relativistic (GR) effects will prohibit the collision during the slow secular evolution leading to it, c. a small fraction of parameter space $\sim(R_{\rm WD}/a)^{1/2}$ available for producing the required extreme eccentricity $1-e\sim R_{\rm WD}/a$ according to the standard Kozai-Lidov theory.  

We show that all of these challenges are overcome in moderately hierarchical triple systems (outer pericenter separations of $r_{p,\rm out}/a\sim 3-10$). The first challenge is overcome with time and persistence-- as we show in \sref{sec:time}, the phase-space neighborhood of zero angular momentum $J=0$ (corresponding to $e\ra1$) is stochastically scanned and the WDs are likely to collide after a sufficient amount of orbits, $N\sim a/R_{\rm WD}\sim 10^6(a/30\rm AU)$. For separations $a<300 \rm AU$, they do so in less than $5\Gyr$. The second challenge is overcome with a single-punch knockout --  the perturber is close enough to significantly change the angular momentum of the eccentric binary at the last apocenter prior to the collision, leading to a jump in pericenter separation $r_p$ from several $R_{\rm WD}$  (far enough to avoid GR and tides) to $r_p<2R_{\rm WD}$ (a head-on collision). This is shown in \sref{sec:rpout_a_cond}. Finally, the third challenge is overcome due to  the breakdown of the approximations underlying the Kozai-Lidov theory, which allows a finite range of parameter space to access $J=0$ ($e=1$)\cite{Ford00,Naoz11,Katz11,Lithwick11,Naoz11b,Bode12}. This permits a relatively broad range of initial conditions that can produce the required extreme eccentricity. This is discussed in \sref{sec:secular}. 
  
We perform $\sim$ ten thousand three-body simulations, each with over $10^6$ orbits, to study the collision fraction in moderately hierarchical triple systems. In order to avoid the limitations of Kozai-Lidov theory, these are done by direct numerical three-body integration. Given the observed strikingly narrow mass function of WDs (e.g. \cite{Liebert05}), we focus on equal mass WD-WD binaries.  We find that a few percent of systems with $r_{p,\rm out}/a=3-10$ experience collisions within $5\Gyr$, as shown in Fig. \ref{fig:N_ain}. We restrict to ``clean'' collisions in which all the pericenter passages preceding the collision are sufficiently distant so that tidal and GR effects are negligible. By numerically including GR and tidal precession in many of the runs, their negligible effect on clean collisions is confirmed. 

The few-percent fraction, which is $\sim 30$ times higher than naively expected $10^{-3}$ from standard Kozai-Lidov theory, is shown to be due to the breakdown of the double-averaging approximation\cite{Bode12} as discussed in \sref{sec:secular}. The breakdown of the quadrupole approximation discussed in \cite{Ford00,Touma09,Naoz11,Katz11,Lithwick11,Naoz11b} does not play a role in this study due to the zero octupole in the equal mass inner binaries and the breakdown of the test particle approximation (emphasized by \cite{Naoz11}) is shown to be irrelevant. 

The possible implications for SNe Ia are 
briefly discussed in \sref{sec:Discussion} with emphasis on 
the remaining challenges including the significance of stellar 
evolution which is not taken into account in this analysis. Finally, we note
that collisions involving other astronomical objects in triple systems, including main sequence stars, neutron stars and black holes, are also  likely much more common than previously believed due to the same considerations leading to WD-WD collisions.

\section{Overcoming the challenges for collisions}\label{sec:Challenges}
In this section we demonstrate analytically that the two main challenges faced by direct collisions in triple systems are avoided in moderately hierarchical configurations. 

\subsection{``Clean'' collisions achieved by single orbit jumps for $3\lesssim r_{p\rm out}/a\lesssim 10$}\label{sec:rpout_a_cond}
%:sec:rpout_a_cond
If close approaches occur during the evolution prior to a collision, the stars may be tidally disrupted or the energy can be dissipated, possibly prohibiting the collision. This challenge is avoided if all pericenter passages preceding the collision are sufficiently distant.  

For simplicity we conservatively restrict to ``clean collisions'', which are close approaches that satisfy:
\begin{itemize}
\item Collision: the two WDs approach a collision distance $R_{\rm col}=2R_{\rm WD}=2\times 10^9\cm$ with a Keplerian pericenter of 
\begin{equation}\label{eq:collision}
%:eq:rpk
\rm{collision:}~~~r_{p}<R_{\rm col}=2\times 10^9\cm,
\end{equation}
where $r_{p}$ is the Keplerian pericenter in the limit $e\ra1$,
\begin{equation}\label{eq:rpK}
r_{p}\equiv \frac{(J/\mu)^2}{2G\Minp},
\end{equation}
where $\mu=m_1m_2/(m1+m2)$ is the reduced mass, and $J$ is the angular momentum of the inner orbit ($J/\mu$ is the specific angular momentum). 
\item ``clean'':  there is no approach prior to the collision that is sufficiently close to allow significant dissipation. This requirement is implemented by demanding that all earlier approaches are greater than a conservative dissipation scale $R_{\rm dissip}=2R_{\rm col}=4\times 10^9\cm$, justified in \sref{sec:GRTide}.
\begin{equation}\label{eq:clean}
%:eq:clean
\rm{clean: }~~~r_p(t<t_{\rm col})>R_{\rm dissip}=4\times 10^9\cm,
\end{equation} 
where $t_{\rm col}$ is the time it takes to reach the collision.
In particular, this implies that the pericenter must change from $r_p>R_{\rm dissip}$ to $r_p<R_{\rm col}$ in one orbit.
\end{itemize}
%In the numerical examples we adopt $R_{\rm dissip}=2R_{\rm col}$.

Consider next the requirement that a significant change in pericenter separations occurs between successive pericenter passages. In terms of the change in the (specific) angular momentum, this can be quantified as 
\begin{equation}\label{eq:DJ_req}
%:eq:DJ_req
\inv{\mu}\Delta J_{\rm orb}\gtrsim\sqrt{2G\Minp R_{\rm dissip}}
\end{equation}
where the right hand side is the specific angular momentum of an orbit at the boundary allowed by dissipation, $r_p=R_{\rm dissip}$. In the quadrupole approximation, the integrated change in angular momentum achieved between two successive pericenter passages is 
\begin{equation}\label{eq:DJ}
%:eq:DJ
\inv{\mu}\Delta J_{\rm orb}=P\frac{15Gm_3a^2}{2r_{\rm out}^3}(\hat e\cdot\hat r_{\rm out}) \hat e\times \hat r_{\rm out}\end{equation}
where $r_{\rm out}$ is the distance between the center of mass of the binary and the perturber, $P$ is the inner orbital period, the relevant $e\ra 1$ was assumed, and the motion of the perturber was neglected. Most of this angular momentum change is obtained in the vicinity of the apocenter. 
The maximal possible change is obtained when $\hat e\cdot\hat r_{\rm out}=1/\sqrt{2}$ implying an upper limit for the angular momentum change of
\begin{equation}\label{eq:DJ_max}
%:eq:DJ_max
\inv{\mu}\abs{\Delta J}_{\rm orb, max}=\frac{15Gm_3a^2P}{4r_{\rm out}^3}.
\end{equation}
We numerically verified the validity of this expression. 
 
By substituting Eq. \eqref{eq:DJ_max} in Eq. \eqref{eq:DJ_req}, the following requirement is obtained on the perturber separation,

\begin{align}\label{eq:j_kick_req}
%:eq:j_kick_req
&\frac{r_{\rm out}}{a}<\ADJ\left(\frac{m_3}{\Min}\right)^{1/3}\left(\frac{a}{R_{\rm dissip}}\right)^{1/6}\cr
&\approx 14 \left(\frac{m_3}{\Min}\right)^{1/3}\left(\frac{a}{30{\rm AU}}\right)^{1/6}\left(\frac{R_{\rm dissip}}{4\times 10^9\rm cm}\right)^{-1/6}.\cr
\end{align}

Eq. \eqref{eq:j_kick_req} shows that clean collisions can only occur for moderately hierarchical systems which allow for significant angular momentum kicks between successive pericenter passages.

\subsection{For $a/R_{\rm WD}\sim 10^6$, collisions are stochastically achieved over $10^6$ orbits}\label{sec:time}
White dwarf binaries can overcome the huge ratio between their typical separation $r\sim a$ and their sizes $R_{\rm WD}$ to achieve collision because they have ample time (orbits) available. How much time is required to achieve a collision?
 
Given the large angular momentum kicks between successive pericenter passages, it is natural to expect that after a large amount of orbits, the (equal energy) phase space is stochastically scanned and roughly uniformly covered by the pericenter values. 

Such a uniform phase space distribution implies a uniform distribution of the  angular momentum squared $J^2$ which corresponds to a uniform distribution of pericenters \begin{equation}\label{eq:dNdrp}
%:eq:dNdrp
\frac{dN}{dr_p}\approx \frac{2}{a}.
\end{equation} 

This is shown to agree with a numerical example in section \sref{sec:example}, figure \ref{fig:example_dNdr}.   

After $N\sim a/2R_{\rm col}$ orbits, a collision is expected. The time to reach collision, $t_{\rm col}$ is expected to follow a Poisson distribution with a mean value given by
\begin{align}\label{eq:T_col}
%:eq:T_col
&t_{\rm col, exp}=\frac{a}{2R_{\rm col}}P\sim 2\times 10^7\left(\frac{a}{30\rm AU}\right)^{5/2}\yr\cr
&\times\left(\frac{m_1+m_2}{M_{\odot}}\right)^{-1/2}\left(\frac{R_{\rm col}}{2\times 10^9\cm}\right)^{-1}.\cr
\end{align}
This distribution is shown to roughly agree with numerical results in \sref{sec:example} (figure \ref{fig:dNdtcol}, including a factor of $2$ explained in \sref{sec:Rdisp_sup}).

It follows from Eq. \eqref{eq:T_col} that the available $\Gyr$ time scale is more than enough for triples with typical inner separations $a\sim 30\rm AU$ to collide. In fact in order for a collision to occur on timescale shorter than $\sim 5\Gyr$ the semi-major axis may be as large as 
\begin{align}\label{eq:a_lim}
%:eq:a_lim
&a\lesssim 300\left(\frac{m_1+m_2}{M_{\odot}}\right)^{1/5}\left(\frac{R_{\rm col}}{2\times 10^9 \cm}\right)^{2/5}\rm AU\,\cr
\end{align}
which is satisfied by most systems.

\section{Few percent of systems collide in direct numerical integrations}\label{sec:Numeric}
%:sec:Numeric
We performed $\sim$ten thousand numerical 3-body integrations using a C-based code that was written for this purpose. Many of the runs include corrections due to GR precession and equilibrium (non-dissipative) tidal deformations by the addition of appropriate (effective) radial attraction forces (Eqs. \ref{eq:GR_U} and \ref{eq:Tide_U}). For the tidal force, conservatively large apsidal constants $k_1=k_2=1$ (Love numbers $k_{L,1}=k_{L,2}=2$) are adopted.  In all runs, the short pericenter passages are resolved by using an adaptive time step. In most of the runs, a symplectic, second order, Preto-Tremaine-Mikkola-Tanikawa\cite{Preto99,Mikkola99a,Mikkola99b} (PTMT) integrator is used with an adaptive time step $\Delta t\propto \abs{U}^{-3/2}$ where $U$ is the total potential energy (see \sref{sec:integrators} for more details). For sufficiently weak perturbers $r_{p,\rm out}\gtrsim 5a$, an additional (non-symplectic) integrator is used for comparison, which incorporates a Wisdom-Holman (WH) \cite{Wisdom91} operator splitting with a high order (8-6-4) coefficient set taken from \cite{Blanes12}, and an adaptive time step $\Delta t\propto r^{3/2}$ where $r$ is the distance between $m_1$ and $m_2$. More details are provided in \sref{sec:numint}. 
%Energy is conserved in all runs to an accuracy better than $\abs{\Delta E}/E_K<0.01$ at pericenter with the vast majority of runs having $\abs{\Delta E}/E_K<10^{-4}$ where $E_K$ is the instantaneous kinetic energy.

In all runs, the WDs have masses $m_1=m_2=0.5$ and radii $R_1=R_2=10^9\cm$ while the perturber's mass is either $m_3=0.5 M_{\odot}$ (most runs) or  $m_3=1 M_{\odot}$. 
Parameters are provided in the Jacobi coordinates (inner orbit: $m_1$ and $m_2$, outer orbit: center of mass of the $m_1$-$m_2$ system and $m_3$).  

We note that most of the long term integrations presented here are not strictly converged individually. This is due to the fact that in many of the systems considered, the relevant time scales are longer than the Lyapunov time scale by orders of magnitude, inhibiting the possibility of convergence. In such cases however, strict convergence is not interesting given that practically any arbitrarily small deviation would lead to significant changes. In these cases, the statistical properties of the evolution are the relevant quantities and these are verified to be converged by using different time resolutions and integration schemes. One important exception, which is strictly converged, is the example presented in fig \ref{fig:example_t_r_a}. In this case the collision occurs relatively early on and strict convergence is easily achieved with the high order WH integrator.  

\subsection{A numerical example of a clean collision}
\label{sec:example}
%:sec:example
\begin{figure}[h]
\includegraphics[scale=0.8]{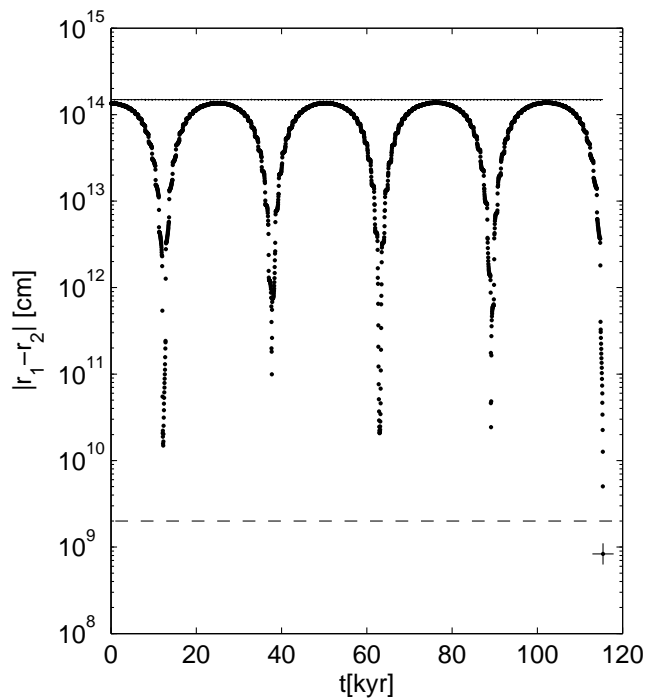}
\caption{Evolution of a triple system that leads to a collision. Shown are the pericenter separations (black dots) and semi-major axis $a$ (solid line)  as a function of time for a $m_1=m_2=0.5 M_{\odot}$ white dwarf binary ($Gm_1=Gm_2=6.637\times 10^{25}\cm^3\sec^{-2}, R_1=R_2=10^9\cm$) with initial $a$=10AU and $e=0.1$ that is orbited by an equal mass, $m_3=0.5 M_{\odot}$ perturber with $a_{\rm out}=100$AU and  $e_{\rm out}=0.5$ (pericenter $r_{p,\rm out}=5a=50$AU). The eccentricity vectors of the orbits are initially aligned while the angular momenta have an initial mutual inclination of $i=98^{\circ}$ (values in the range $95^{\circ}-100^{\circ}$ produce similar evolution). The initial true anomaly of the perturber is $f_{\rm out}=0$ while that of the binary is exactly $f=10^{-4}\rm rad$, chosen by trial and error to achieve an early collision for demonstration purposes (see fig \ref{fig:dNdtcol} for the collision time distribution of similar systems). General relativistic precession and equilibrium (non-dissipative) tidal attraction of the WD binary are included (see Eqs. \ref{eq:GR_U},\ref{eq:Tide_U}), the latter with apsidal constants $k_1=k_2=1$. The values of each pericenter are numerically converged.
A collision occurs at $t=115.4\,$kyr.  The WDs approach the collision with a Keplerian pericenter $r_{p}=8.34\times 10^8\cm$ (Eq. \ref{eq:rpK}, plotted as a plus symbol). This is deep within the collision separation of $R_1+R_2=2\times 10^9\cm$ (plotted as a horizontal dashed line). This is a ``clean'' collision since the closest approach in the evolution prior to the collision was $5.0\times 10^9\cm\approx 5 R_{\rm WD}>R_{\rm dissip}=4 R_{\rm WD}$ (see Eq. \ref{eq:clean}). \label{fig:example_t_r_a}}
%:fig:example_t_r_a
\end{figure}

\begin{figure}[h]
\includegraphics[scale=0.8]{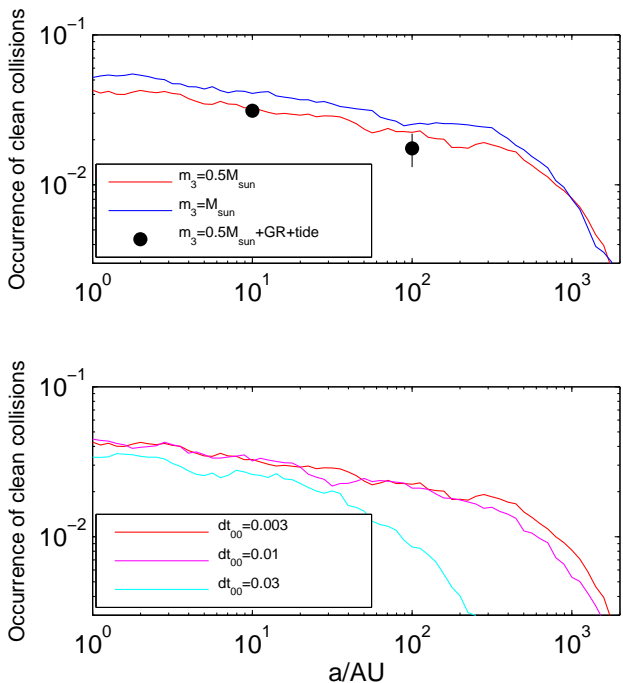}
\caption{Top panel: Fraction of moderately hierarchical systems ($3<r_{p,\rm out}/a<10$) that experience clean collisions within $5 \Gyr$. The systems considered include a WD binary with $m_1=m_2=0.5M_{\odot}$, $R_1=R_2=10^9\cm$ and semi-major axis spanning $a=1-2000 \rm AU$ orbited by a perturber with mass  $m_3=0.5M_{\odot}$ (red solid line, black dots) or $m_3=M_{\odot}$ (blue solid line). The pericenter of the perturber's orbit is assumed to be distributed uniformly in $\log r_{p,\rm out}$ in the range $3a<r_{p,\rm out}<10a$ and both orbits have a uniform distribution of eccentricities in the range $e=0-0.9$. Results based on Newtonian (no GR or tide) ensembles are shown in red (blue) solid lines for perturber mass $m_3=0.5M_{\odot}$ ($m_3=M_{\odot}$). These ensembles B3-B10 (G2-G11), described in table \ref{tab:Simulations}, involve dimensionless distances and times, and the results are computed for each value of $a$ by appropriate scalings. 
The Newtonian results for $m_3=0.5M_{\odot}$(red solid line) are compared with non-Newtoninan results (black dots with statistical error-bars), including GR and tidal precession based on ensembles A3-A10 ($a=10$AU) and F ($a=100$AU).  The GR and tides are implemented using Eqs. \ref{eq:GR_U},\ref{eq:Tide_U} with apsidal constants $k_1=k_2=1$. As can be seen, a few percent of moderately hierarchical  systems experience clean collisions, with GR and tides having a negligible effect. This is the main result of the paper.  
Lower panel: Numerical convergence test. The results for the ensembles B3-B10 (red solid line, $m_3=0.5M_{\odot}$, $dt_{00}=0.003$, see \sref{sec:integrators}) are compared with two additional runs with larger time steps, $dt_{00}=0.01$ (magenta) and $dt_{00}=0.03$ (cyan). As can be seen, the result is well converged.
\label{fig:N_ain}}
%:fig:N_ain
\end{figure}
It is instructive to first consider an example of a 3-body system that experiences a collision. Figure \ref{fig:example_t_r_a} shows the evolution of such a system in which the WD binary has a semi-major axis of $a=10\rm AU$, a perturber with mass $m_3=0.5 M_{\odot}$ and semi-major axis of $a_{\rm out}=100 \rm AU$ and an initial mutual inclination of  $98^{\circ}$ (the exact parameters are given in the caption).  General relativistic precession and equilibrium (non-dissipative) tidal attraction of the WD binary are included (a similar clean collision occurs after $3.85\times 10^5\yr$ in a Newtonian integration with no GR and tides with an initial inner true anomaly of $f=-0.02\rm rad$ and identical parameters otherwise)
. While the semi-major axis of the WD binary is almost constant (solid line), the pericenter varies significantly due to the presence of the perturber. The two white dwarfs experience a head-on collision (approaching with a Keplerian pericenter of $r_p=0.83 R_1$, shown as a '+' symbol)  at $t_{\rm col}=115.4$kyr. 

The slow (20 kyr scale), periodic variations in the pericenter are the Kozai-Lidov \cite{Lidov62,Kozai62} oscillations mentioned in \sref{sec:Introduction} that are known to occur in highly inclined triple systems. These occur on time scales much longer than the inner period ($P\sim30 {\rm yr}$) and the outer period ($P_{\rm out}\sim 1$kyr) as expected. In contrast to this slow periodic evolution, the values of the pericenter separations can change significantly and stochastically between successive pericenter passages. In particular, the pericenter passage that occurred one orbital period prior to the collision has $r_p = 5.0\times 10^9\cm\approx 5 R_{\rm WD}$ while at collision $r_p=1 R_{\rm WD}$.  Throughout the evolution prior to the collision, the white dwarfs are always separated by distances larger than $5 R_{\rm WD}$ satisfying the clean collision criteria and implying that dissipation (tidal dissipation and gravitational radiation) could not stop the collision.

\begin{figure}[h]
\includegraphics[scale=0.8]{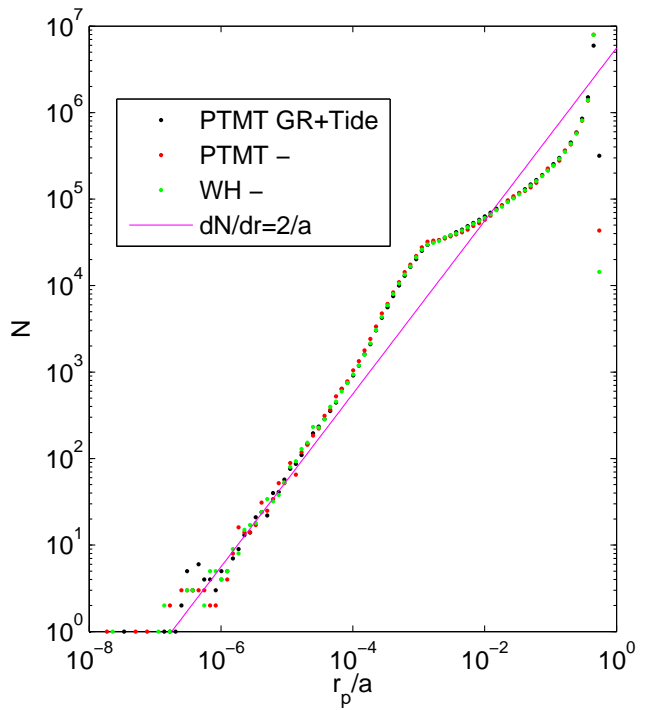}
\caption{Distribution of pericenter separations in the example shown in figure \ref{fig:example_t_r_a}. In order to increase the statistics, the pericenter approaches for hundreds of systems with randomly chosen mean anomalies (inner and outer orbits) with all other parameters being identical, were collected.  
The results from different integrators agree (PTMT  including GR+tide in black dots, Newtonian PTMT in red dots and Newtonian WH in green dots), confirming the convergence of the result.
At low pericenters, the distribution is uniform and agrees with Eq. \eqref{eq:dNdrp} (magenta), unaffected by GR and tides. 
\label{fig:example_dNdr}}
%:fig:example_dNdr
\end{figure}

 The distribution of pericenter passages of this example is presented in figure \ref{fig:example_dNdr} and shown to agree with the expected distribution in Eq.\eqref{eq:dNdrp} to a good accuracy at close approaches. This comparison is achieved by accumulating the results of hundreds of integrations in which the inner and outer true anomalies are randomly varied while leaving the other initial conditions unchanged.

\subsection{Numerical evaluation of the chance of a clean collision in moderately hierarchical triples}\label{sec:NumEns} 
How much fine tuning is required to achieve clean collisions? We next address this question by brute force numerical experiments which are summarized in table \ref{tab:Simulations}.

Two large sets of ensembles with varying values of $r_{p,\rm out}/a$ are used for estimating the collision fraction. Both have  $m_1=m_2=m_3=0.5M_{\odot}$, and use the PTMT with a step size of $dt_{00}=0.003$ (see \sref{sec:integrators} and in particular,  Eqs. \ref{eq:dt0}, \ref{eq:dt00}).  
The two sets are: 1. a set of non-Newtonian ensembles (with GR and tidal preccesion)  having a semi-major axis of $10\rm AU$ (A2-A10)  2. a set of Newtonian ensembles (no GR or tides) with dimensionless distance and the results are scaled to estimate the collision fraction at $a=1-3000\rm AU$  (B2-B10).  

There are three additional sets of ensembles which are used to check the robustness of the result. These are 1. A large set of Newtonian ensembles with a perturber mass $m_3=1M_{\odot}$ (G1-G11). 2. A non-Newtonian ensemble with $a=100\rm AU$ (F). 3. a set of Newtonian ensembles which use the WH integrator with a step size of $dt_0=0.1$    
 (see \sref{sec:integrators}, Eq. \ref{eq:dt0WH}, C1-C3).

\emph{Initial conditions}
The ratio $r_{p,\rm out}/a$ is either fixed to a given value or randomly chosen from a log-uniform distribution in the range $3-10$. The eccentricities of both orbits are randomly chosen from a uniform distribution in the range $0<e<0.9$ and the semi-major axis of the outer orbit is then calculated from the known pericenter and eccentricity.  The mean anomalies are chosen randomly and the orbital orientations are randomly chosen from an isotropic distribution. 

\begin{figure}[h]
\includegraphics[scale=0.84]{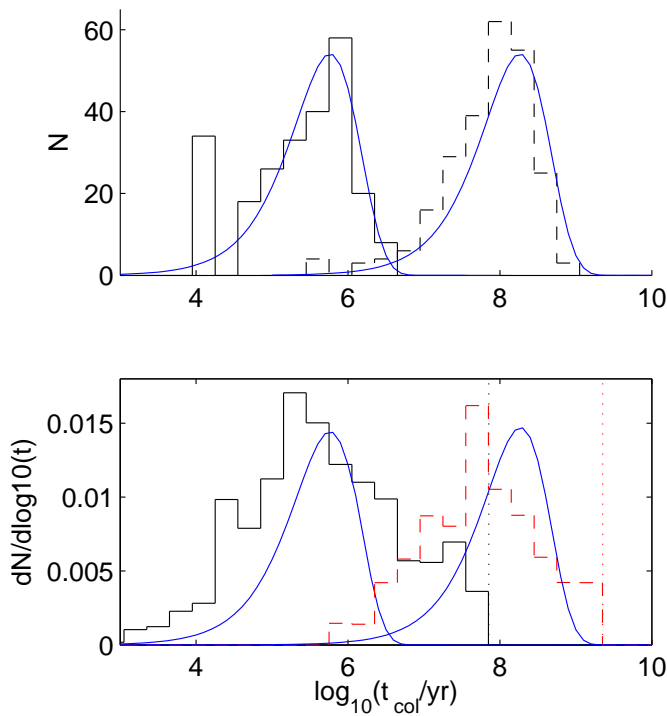}
\caption{Distribution of times to reach collision $t_{\rm col}$. Upper panel: $t_{\rm col}$ distribution in ensembles with parameters identical to the example presented in figure \ref{fig:example_t_r_a},  except for mean anomalies which are uniformly distributed. GR and tidal precession are included in the integrations. Distributions are presented for $a=10$AU (solid black) and for $a=100$AU (dashed black). The blue solid lines in both panels are the theoretical poisson distribution $dN/dt\propto t\exp(-t/t_{0})$ with $t_0=5.9\times 10^5\yr$ ($1.9\times 10^8\yr$) calculated by using Eq. \eqref{eq:T_ccol} for $a=10\rm{AU} (100\rm AU)$. 
Bottom panel: $t_{\rm col}$ distribution for the ensembles with $3<r_{p,\rm out}/a<10$ and $m_3=0.5M_{\odot}$. Distributions are presented for $a=10\rm AU$ (black solid, including GR and tidal precession, based on ensemble A3-A10) and for $a=100\rm AU$ (red dashed, Newtonian, based on ensemble B3-B11). The maximal integration times $t_{\max}$ are shown as dotted lines. Any collision occurring beyond $t_{\rm max}$ is missed. As indicated by the figure, the amount of collisions missed by the simulations is insignificant.\label{fig:dNdtcol}}
%:fig:dNdtcol
\end{figure}

\emph{Stopping conditions}
The evolution is stopped if one of the following conditions are met:
\begin{itemize}
\item one of the stars becomes unbound and is ejected from the system;  
\item the white dwarfs become closer than  $R_{dissip}=4\times 10^9\cm$. In this case, the angular momentum of the orbit near the collision is recorded and the Keplerian pericenter (Eq. \ref{eq:rpK}) is calculated. If this pericenter is smaller than $R_{\rm col}=2\times 10^9\cm$, the system is considered to have experienced a clean collision;
\item 5 Gyr has passed;
\item A predetermined maximal integration time $t_{\max}/P\approx 2\times 10^6$ WD orbits is reached. Note that the results of Newtonian simulations  (where GR and tidal precession are not included) can be scaled in distance and time, and $t_{\max}$ may be smaller than $5\Gyr$.  This implies that the occurrence reported here should be considered as a lower limit. The true occurrence is likely not much higher, as indicated by the distributions of collision time in figure \ref{fig:dNdtcol} and the discussion in \ref{sec:Rdisp_sup} and figure \ref{fig:tmax}.
\end{itemize}

\begin{figure}[h]
\includegraphics[scale=0.8]{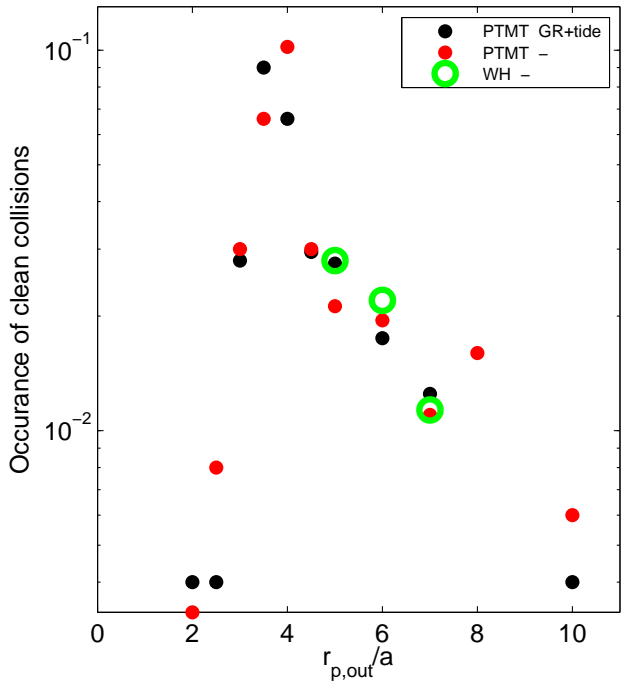}
\caption{Fraction of systems that experience collisions as a function of $r_{p,\rm out}/a$. The Newtonian (red) and non-Newtonian (black, with GR +tide) are based on the same ensembles with identical color coding and the.  The results, which are computed with the PTMT integrator, are shown to agree with those using the WH integrator (green circles, see \sref{sec:integrators}. A few simulations  the Wisdom-Holman integrator (WH, see \sref{sec:integrators}) are shown as green circles. 
\label{fig:N_rpout_ain}}
%:fig:N_rpout_ain
\end{figure}

\emph{Results}
The fraction of systems that experience clean collisions as a function of the WD binary semi-major axis $a$, and the pericenter of the perturber $r_{p,\rm out}$ (normalized to $a$)  is shown in figures \ref{fig:N_ain} and \ref{fig:N_rpout_ain}, respectively. The fraction in figure \ref{fig:N_ain}  is calculated assuming a log-uniform distribution of perturber pericenters in the range $r_{p,\rm out}/a=3-10$. 

As can be seen, a few percent of such systems experience clean collisions within $5 \Gyr$ over a broad range of semi-major axis values. This is the main result of this paper.

The fraction is significant over a limited range of values of $r_{p,\rm out}/a$. At the lower end, the systems eject one of the stars before they have time to collide, while at the higher end the decrease in the fraction is primarily due to the decreasing range of collision-permitting inclinations (see \sref{sec:secular}).

As can be seen in figures \ref{fig:N_ain}, and \ref{fig:N_rpout_ain}, GR and tidal precession do not appear to modify the collision fraction for this range of pericenters. The distribution of time to reach collision is shown in the bottom panel of figure \ref{fig:dNdtcol}. As can be seen it is very broad, varying by orders of magnitude and in rough agreement with Eq. \eqref{eq:T_ccol}.

\section{The significant fraction of collisions is due to the breakdown of the double-averaging approximation}\label{sec:secular}
In order for collisions to occurr, the region of phase-space near $J=0$ needs to be reached by the inner orbit during the evolution. As explained in \sref{sec:Introduction}, for distant perturbers for which standard Kozai-Lidov theory applies, this region is often avoided by the long term evolution unless the initial inclination between the inner and outer orbits is highly tuned ($10^{-3}$ tuning required for the $1-e\sim 10^{-6}$ values required for collisions). We next shortly review the essential arguments of this theory and show that it breaks down in the scenarios discussed here due to the failure of the ``double-averaging'' approximation \cite{Bode12}. This leads to a much broader range of allowable initial inclinations to achieve high eccentricities (practically independent of how high), essential for the high chances of collisions (few percents) reported here. We stress that the evolution for tight systems, $r_{p,\rm out}\lesssim 4 a$, is unlikely to be captured by the simple analysis presented below. 

There are three assumptions commonly attributed to the standard Kozai-Lidov theory:
\begin{enumerate}
\item ``Double averaging'' approximation - the evolution is slow and the equations of motion can be averaged over the rapidly varying mean anomalies of the inner and outer orbits.
\item ``quadrupole'' - Only the leading term (second order in $r/r_{\rm out}$) in the multipole expansion of the interaction potential is kept. 
\item ``Test particle'' - the mass of one of the objects in the inner binary is assumed to be 0.  More precisely, the angular momentum of the inner orbit is much smaller than that of the outer one $J_{\rm out}\gg J_{\rm in}$.
\end{enumerate}
The crux of the Kozai mechanism is the fact that the potential induced by the perturber is axisymmetric when averaged over the outer orbit. This is a straightforward yet surprising outcome of the quadrupole approximation.  

In the test particle approximation (approximation 3), the angular momentum $J_{\rm out}$ of the outer orbit is constant. The axisymmetry implies that the $J_z$ component of the inner angular momentum is conserved where $z$ is chosen along the direction of $J_{\rm out}$. The ``Kozai constant'' is simply a dimensionless costume of $J_z$
\begin{equation}\label{eq:jz}
j_z=\frac{J_z}{J_{\rm circ}}=\sqrt{1-e^2}\cos i.
\end{equation}  
where 
\begin{equation}\label{eq:Jcirc}
J_{\rm circ}=\mu\sqrt{G(m_1+m_2)a},
\end{equation}
$\mu=m_1m_2/(m_1+m_2)$ is the reduced mass and $i$ is the mutual inclination between the angular momenta of the two orbits.

A direct consequence of equation \eqref{eq:jz} is that the magnitude of the inner orbit's angular momentum has a lower limit (its constant $z$ component) and the eccentricity has an upper limit
\begin{equation}\label{eq:emaxjz}
1-e>1-(1-j_z^2)^{1/2}\approx 0.5j_z^2.
\end{equation}
Given that random isotropic initial conditions have a uniform $-1<j_z<1$ distribution, the chance of achieving the extremely high eccentricities $1-e\sim 10^{-6}$ required for collisions is thus very small, $(1-e)^{1/2}\sim 10^{-3}$, and is much lower than the numerically obtained collision fraction shown in fig \ref{fig:N_ain}. We next provide an explanation for this difference, focusing on perturbers with $r_{p,\rm out}/a\sim 5$. 

 It is reasonable to examine the three Kozai-Lidov assumptions given the proximity of the perturber. We next examine each of these and conclude that the failure of the standard Kozai-Lidov theory in describing these systems is due to the breakdown of assumption 1, the double averaging approximation.

 Consider first the implication of relaxing the test particle approximation (e.g. \cite{Naoz11}, assumption 3.). It turns out that this has insignificant effects on the outcome (\cite{Lidov76},  unlike the somewhat confusing discussion in \cite{Naoz11}). The reason is simple- the axisymmetry is unrelated to this assumption. There is no torque along the direction of $J_{\rm out}$ (which is the axis of the symmetry) and the magnitude of $J_{\rm out}$ is conserved ($e_{\rm out}={\rm const}$, \cite{Lidov76}).  Within the double-averaging+quadrupole approximation, the following generalization of $j_z$ is constant,
\begin{align}\label{eq:jzeffdeff}
%:eq:jzeffdeff
&j_{z,\rm eff}\equiv \frac{\abs{J_{\rm tot}}^2-\abs{J_{\rm out}}^2}{2\abs{J_{\rm out}}J_{\rm circ}}\cr
&=j\cdot\hat J_{\rm out}+j^2\frac{J_{\rm circ}}{2\abs{J_{\rm out}}}
\end{align}
where $J_{\rm tot}$ is the total (and always constant) angular momentum of the system,  and $J_{\rm circ}$ is defined in Eq. \eqref{eq:Jcirc}. In the limit $J\ll J_{\rm out}$, $j_{z,\rm eff}\ra j_z$, where $\hat z$ is now the direction of the total angular momentum. 

For isotropic system orientations, $j_{z,\rm eff}$ is (roughly) uniformly distributed and the amount of inclination tuning, required to achieve high eccentricities, is approximately the same as that in the test particle approximation. The only modification in the non-test particle cases is a modest shift in the values of the inclinations required for high eccentricities from being centered on $90^\circ$ in the test particle to slightly higher values (e.g. $\approx 97.5$ in the example considered in \sref{sec:example}) in the non-test particle. The width of the range of allowed inclinations is essentially left unchanged.

The breakdown of either the quadrupole approximation (octupole or higher contribution) or the double-averaging assumptions has been shown to lead to high eccentricities without the need of high inclination tuning (octupole - \cite{Ford00,Touma09,Naoz11,Katz11,Lithwick11,Naoz11b}, double averaging - \cite{Bode12}.

In this paper we focus on triples with equal mass inner binaries for which there is no octupole contribution\footnote{The octupole potential, induced by the perturber, has a cubic dependence on the inner separation vector and is anti-symmetric with respect to mirror- reflections. In the equal mass case, the time spent by the binary at different separations within the  Keplerian orbit is symmetric to reflections perpendicular to the eccentricity vector. This leads to a vanishing orbit averaged contribution.} implying that approximation 2 is valid. We made this choice because WDs have a strikingly narrow mass distribution around $M\approx 0.6 M_{\odot}$ (e.g. \cite{Liebert05}).

\begin{figure}[h]
\includegraphics[scale=0.8]{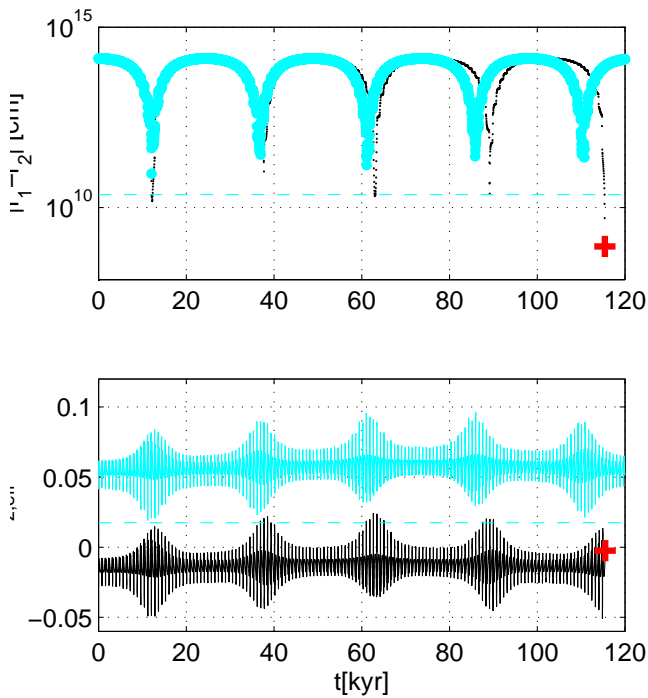}
\caption{Evolution of $j_{z,\rm eff}$ (lower panel, defined in Eq. \eqref{eq:jzeffdeff}) and binary pericenter separations (upper panel) for the high inclination $i=98^{\circ}$ example described in fig \ref{fig:example_t_r_a} (black) as well as a system with slightly lower inclination, $i=94^{\circ}$ which is otherwise identical (cyan). The red plus marks the collision experienced by the higher inclination system. In both systems, $j_{z,\rm eff}$ experiences variations on the outer orbit timescale ($\sim1$kyr) which are enhanced to $\sim 0.1$ at the eccentricity peaks within each Kozai-Lidov cycle ($P_{Kozai-Lidov}\sim$20kyr). For the $i=94^{\circ}$ case, these variations are not sufficient to bring $j_{z,\rm eff}$ to $0$. The minimal value of $j_{z,\rm eff,\min}\approx0.017$ obtained in this example (lower panel, cyan dashed line) sets a lower limit of $r_{\min}=0.5j_{z,\rm eff,\min}^2a\approx 2.2\times 10^{10}\cm$ to the WD-WD separation (upper pannel, cyan dashed line) which prohibits the possibility of a collision. 
\label{fig:example_jzeff_GRWHPT}}
%:fig:example_jzeff_GRWHPT
\end{figure}

As we next show, the double-averaging assumption breaks down and the available range of inclinations for achieving high eccentricities is much larger than that expected in the standard Kozai-Lidov theory.  This is illustrated in figure \ref{fig:example_jzeff_GRWHPT}, where the value of $j_{z,\rm eff}$ from Eq. \eqref{eq:jzeffdeff} for the example studied in \sref{sec:example} is shown. As can be seen, $j_{z,\rm eff}$ experiences variations on the outer orbit timescale ($\sim1$kyr) which is a clear signature of the breakdown of the double-averaging approximation These variations occur because the instantaneous quadrupole potential is not axisymmetric\cite{Bode12}. Note that the amplitude of these variations is biggest during the high eccentricity phases of Kozai-Lidov cycles ($P_{\rm Kozai-Lidov}\sim 20\rm kyr$) due to the larger apocenters where most of the angular momentum exchanges occur. A rough estimate of the variation size can be obtained by considering the angular momentum kick achieved within a fraction of the pericenter passage, say $\Delta t\sim\pi r_{p,\rm out}^{3/2}[G(m_1+m_2+m_3)]^{-1/2}$. Using Eq. \eqref{eq:DJ_max} and assuming that the component of the torque in the direction of the angular momentum is of the scale of the torque's magnitude, the following estimate is obtained for the fluctuations of $j_{\rm z, eff}$ within the outer orbit,
\begin{equation}\label{eq:jz_fluc}
%:eq:jz_fluc
\Delta j_{z,\rm eff,~out~orb}\sim \frac{m_3}{\sqrt{(m_1+m_2)m_{\rm tot}}}\left(\frac{a}{r_{p,\rm out}}\right)^{3/2}.
\end{equation}
Given that $r_{p,\rm out}/a\sim 5$, a fluctuation of order $0.1$ is not surprising. This implies that there is roughly a $0.1/2=5\%$ range of initial $j_{z\rm eff}$ values for which $j_{z\rm eff}$ reaches $0$ during these oscillations. At the high initial inclinations considered we have $\Delta i_0\sim \Delta j_{z\rm eff}$, and so the width of available initial inclinations to access $j_{z,\rm eff}=0$ and reach extreme eccentricity is,
\begin{align}\label{eq:inctun}
%:eq:inctun
\Delta i_0 &\sim\Delta j_{z,\rm eff,~out~orb} \cr
&\sim5^{\circ}\times\frac{m_3}{\sqrt{(m_1+m_2)m_{\rm tot}}}\left(\frac{r_{p,\rm out}}{5a}\right)^{-3/2}.\cr
\end{align}
For the example considered, the available range is $95^{\circ}-100^{\circ}$.  
This allows an increase in available phase space for WD-WD collisions of orders of magnitude as compared to standard Kozai-Lidov theory. The importance of this evolution, that occurs within the period of the outer orbit, was realized by \cite{Bode12} in the context of black-hole triples.

A smaller, long term modulation (on a timescale of $~10$ Kozai-Lidov cycles) observed in fig. \ref{fig:example_jzeff_GRWHPT} is yet to be explained \footnote{We note that these long terms oscillations are periodic and persist for at least $10^6$ orbits.  They are reproduced to good accuracy when integrating the single-orbit-averaged equations to fourth order in $r_{\rm out}/r$ but not to second (the third order is zero for the equal masses considered here)}.

For non-equal mass WDs, the contribution of the octupole allows an additional long term change in $j_{z,\rm eff}$ \cite{Ford00,Touma09,Naoz11,Katz11,Lithwick11,Thompson11,Shappee12}. In these cases even less tuning is required. Given that the mass function of WDs is very narrow however it is likely that most WD binaries have similar masses $\approx 0.6M_{\odot}$. In this paper we conservatively focus on equal mass WDs for which the octupole term vanishes.

\section{Discussion}\label{sec:Discussion}
 In this paper, it was shown that a WD-WD binary which is orbited by a stellar mass perturber with moderate hierarchy ($r_{p,\rm out}/a\sim 3-10$) has a few percent chance of experiencing a head-on collision. This is numerically demonstrated in \sref{sec:Numeric} (in particular see figure \ref{fig:N_ain}) and explained analytically in \sref{sec:Challenges} and\sref{sec:secular}.  It is shown that WD binaries with orbital separations $a<300$AU (Eq. \ref{eq:a_lim}), and initial inclinations with a tuning level of $\Delta i\sim 5^{\circ}$ (or $\sim$ 5\% of isotropic systems, Eq. \ref{eq:inctun}) are likely to experience a collision within $5\Gyr$
%  This relatively broad range of initial conditions is much larger than what is naively expected from the standard Koza-Lidov mechanism. For systems with $r_{p,\rm out}\gtrsim 4a$, this is due to the breakdown of the double-averaging assumption underlying the standard Kozai-Lidov theory \cite{Bode12}. This effect cannot be captured by studies based on double averaging such as \cite{Touma09,Naoz11,Katz11,Lithwick11,Thompson11,Naoz11b}, regardless of the order kept in the perturbation series (quadrupole, octupole etc.). We do not attempt to analyze the more violent evolution at $r_{p,\rm out}\lesssim 4a$. For non-equal-mass WDs (not considered here due to their rarity) the octupole correction may allow a similar or even broader range of allowed inclinations \cite{Naoz11,Katz11,Lithwick11,Shappee12}. 
  The collisions considered here are ``clean'' in the sense that none of the orbits preceding the collision have close encounters with  $r<R_{\rm dissip}=4R_{\rm WD}$ (Eq. \ref{eq:clean}), implying that dissipative (not included in our simulations) and non-dissipative (included) corrections have negligible effects on the resulting collision fraction. The required rapid change of angular momentum, sufficient to significantly change the pericenter between two successive pericenter passages, is achieved due to the moderate hierarchy considered $r_{p,\rm out}/a\lesssim10$ (see Eq. \ref{eq:j_kick_req} and figure \ref{fig:example_t_r_a}).

Such collisions are likely to lead to type Ia SN explosions  \cite{Rosswog09,Raskin09,Raskin10,Hawley12}. Given that the SN Ia rate is roughly $100$ times smaller than that of the WD formation \footnote{e.g.: star formation rate of $3\times10^{-2} M_{\odot}\yr^{-1}\Mpc^{-3}$, WD fraction of $0.1$ WDs per solar mass formed, and SN Ia rate of $3\times 10^{-5}\yr^{-1}\Mpc^{-3}$}, WD-WD collisions may be as common as SN Ia if a significant fraction $\gtrsim 30\%$ of white dwarfs are in such triples. Given the uncertainties in the distribution of multiple systems in the various relevant environments, it is possible that  some or all SNe Ia occur in such collisions. A direct collision of two WDs has a unique gravitational wave signature that will allow in the future a definitive way to test this exciting possibility (e.g. \cite{Loren10}).

The main open question is whether or not a sufficient fraction of WDs spend a sufficient amount of time in such triples. In particular, the stellar evolution leading to the formation of WD binaries, which is not considered in this paper, may play an important role. Specifically, the triple scenario faces the following challenge: If the systems are sufficiently active to allow the WDs to collide, why don't the progenitor stars in the binary collide or have strong encounters at earlier evolutionary stages, when they have much larger radii (e.g. \cite{Thompson11,Shappee12})? While this is a serious concern which we do not intend to resolve in this paper, we note the following: most triple system configurations are stable and do not experience close approaches. The triple configurations may change significantly at the latest stages of mass loss or much later due to interactions with passing stars or with the local tidal field. If such changes are sufficiently violent to ``reset'' the configuration after the two stars have become WDs, the fraction of triples that avoid contact throughout the stellar evolution but collide at a later stage may be similar or even larger than the fractions reported here.
A detailed characterization of the multiplicity properties of B, A, and F-type stars combined with a detailed modeling of the evolution is crucial to better estimate the occurrence of such collisions.    

A second open question is how exactly would such events look (in terms of spectra and light curves).  While a collision is likely to lead to an explosion, the amount of Nickel 56 produced is quite uncertain \cite{Rosswog09,Raskin09,Raskin10,Hawley12} and significant additional work is required to permit a useful comparison with SNe Ia observations.

Finally we note that the considerations derived here are applicable to collisions involving a large variety of other astronomical objects in triple systems, including main sequence stars, neutron stars,  black holes, planets and combinations of these with each other and with WDs. We expect that collisions and extremely close encounters among such objects are much more common than previously believed and are likely to lead to a rich variety of exciting observational outcomes. Probably the simplest and most common collisions  involve main-sequence (MS) stars in MS-MS or MS-WD collisions, which may result in observable transients (e.g. \cite{Shara86,Soker06}). Indeed, for a collision distance of $R_{\rm col}\sim 2R_{\rm sun}$, Eq. \eqref{eq:a_lim} implies that such collisions may occur for inner-binary separations $a<1500$AU, satisfied by the majority of triples in the field.

\acknowledgements We thank Andy Gould and Scott Tremaine for a careful reading of the manuscript and for useful comments. We thank Jose Prieto, Aristotle Socrates, Todd Thompson, Ben Shappee, Ehud Nakar and Doron Kushnir for useful discussions. B.K is supported by NASA through Einstein Postdoctoral Fellowship awarded by the Chandra X-ray Center, which is operated by the Smithsonian Astrophysical Observatory for NASA under contract NAS8-03060. S.D. is supported through a Ralph E. and Doris M. Hansmann Membership at the IAS.

\clearpage
\appendix

\section{Non-newtonian corrections are negligible for $R_{\rm dissip}=4\times 10^9$}\label{sec:GRTide}
In this section, order of magnitude estimates are derived for the influence of non-Newtonian corrections and are shown to be negligible in the clean collision scenarios studied here.

\subsection{Secular precession due to tidal deformation and General relativistic corrections}
Fast apsidal precession suppresses the changes in the magnitude of an orbit's angular momentum by averaging out non-axisymetric components in the potential which are required to apply a torque in the direction of the angular momentum. As illustrated by our numerical results, precession due to equilibrium tides and GR has negligible consequence. 

In order for the exchange of angular momentum to be suppressed, significant precession is required on the time scale of angular momentum change $J/\tau$ where $\tau$ is the typical torque.  For the clean collisions considered in this paper, the time scale for significant change in the angular momentum is shorter than a single orbit. Within one orbit the amount of GR precession is $\Delta \om\sim v^2/c^2$  while that of the equilibrium tide is  $\Delta \om \sim k(r_p/R_{\rm WD})^6$. Both of these are negligible at the closest approaches considered $r>4R_{\rm WD}$. Note that during the collision approach, the attractive force of the tidal deformation can only (slightly) decrease the impact parameter while the GR corrections are negligible.

\subsection{Dissipative corrections}
\emph{Tidal dissipation}
The tidal energy dissipated in each WD in one orbit is parametrized using the tidal quality factor $Q$ and given by
\begin{equation}
\Delta E=k_1Q^{-1}\frac{Gm_2^2R_1^5}{r_p^6}
\end{equation}
where the dissipation occurs in $m_1$ due to the tidal bulge raised on it by the tidal field imposed by $m_2$. Such energy loss may have a significant effect on the evolution if it is comparable to the (absolute value of the) orbital energy $E_{\rm orb}=Gm_1m_2/(2a)$. Given that the occurrence of passages with separations $r_p$ is proportional to $r_p^1$ while the energy lost is suppressed by $r_p^{\alpha}$ with $\alpha$ much larger than unity (the exact value of $\alpha$ depends on $Q$ and the dissipation physics) the dominant contribution would come from the single closest approach before the collision which is restricted to $r_p>R_{\rm dissip}=4R_{\rm WD}$. 
%We thus have: 
%\begin{equation}
%\frac{\Delta E}{E_{\rm orb}}=k_1Q^{-1}\frac{m_2}{m_1}\left(\frac{R_1}{r_{p,\rm in}}\right)^6\frac{a_{\rm in}}{R_1}
%\end{equation}
%
In order that the nearest passage does not change the orbital energy considerably it is required that 
\begin{align}
&Q>2\frac{a}{R_1}k_1\frac{m_1}{m_2}\left(\frac{R_1}{R_{\rm dissip}}\right)^6\sim 200k_1(a/30AU)\cr
\end{align}

In the extremely unlikely case that $Q<200k_1$ (typical estimates of $>10^6$ are given e.g. \cite{Piro}), a modest suppression would be required in the clean collision rate.

\emph{Gravitational wave emission}
At high eccentricities, the gravitational energy radiated per orbit \cite{Peters64} can be conveniently expressed as 
\begin{equation}
\Delta E\approx-16.0\frac{m_1m_2}{(m_1+m_2)^2}\left(\frac{v_p}{c}\right)^{5}E_p
%\approx -2\times 10^{-8}\frac{\mu}{M}\left(\frac{M}{M_{\odot}}\frac{0.01R_{\odot}}{r_p}\right)^{2.5}
\end{equation}
where $v_p=(G(m_1+m_2)/r_p)^{1/2}$  and $E_p=Gm_1m_2/r_p$ are respectively the velocity and gravitational potential energy at pericenter. 
At the closest approaches allowed by our considerations, $r_p=4\times 10^{9}\cm$, the pericenter velocity is about $v_p\sim 1800\km/s\sim 0.006c$. This implies an emission per orbit of $\Delta E\sim 10^{-11}E_p$ which is $10^{-4}(a/30AU)$ smaller than the orbital energy  $E_{\rm orb}=r_p/(2a)E_p\sim 10^{-7}(a/30AU)^{-1}E_p$ and can be neglected.

\subsection{Rate of clean collisions is insensitive to $R_{\rm dissip}$ with a suppression $\propto 1/R_{\rm dissip}$. No need to integrate more than $10^6$ orbits at all $a$}\label{sec:Rdisp_sup}
\begin{figure}[h]
\includegraphics[scale=0.8]{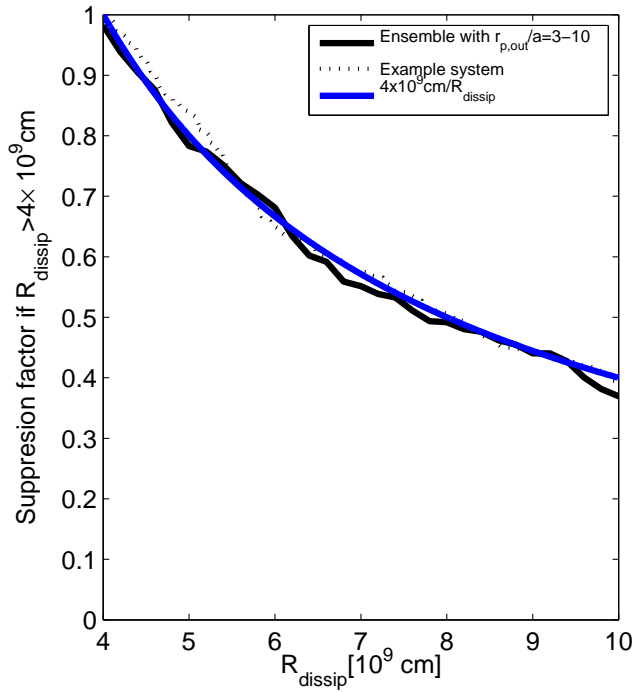}
\caption{Suppression of the fraction of collisions as a function of $R_{\rm dissip}$ in case $R_{\rm dissip}>4\times 10^9\cm$. The black solid line is the suppression found by varying the $R_{\rm dissip}$ in the analysis of the numeric ensemble studied above (the ensemble A3-A10 with a=10AU which includes GR and Tidal precession was used). The dashed solid line is from the example presented in figs \ref{fig:example_t_r_a},\ref{fig:example_dNdr} and the upper panel of \ref{fig:dNdtcol}. The blue solid line is the theoretical expectation for a poisson distribution of pericenters given by Eq. \eqref{eq:Rdissip_sup}.
\label{fig:dNdRdissip}}
%:fig:dNdRdissip
\end{figure}

%:sec:Rdisp_sup
Consider the implication of the requirement that all previous approaches satisfy $r_p>R_{\rm dissip}$. Assuming that $r_p$ is uniformly distributed, the chance that a collision occurs during the first dissipative encounter ($r_p<R_{\rm dissip}$)  is $R_{\rm col}/R_{\rm dissip}$. This implies that the fraction of clean collisions is
\begin{equation}\label{eq:Rdissip_sup}
%:eq:Rdissip_sup
\frac{ \rm clean~collisions}{\rm collisions}=\frac{R_{\rm col}}{R_{\rm dissip}}.
\end{equation}
As can be seen in fig \ref{fig:dNdRdissip}, Eq. \eqref{eq:Rdissip_sup} agrees with the numerical results to a high accuracy.

These arguments imply that clean collisions typically occur earlier than all collisions. In fact, the time distribution of clean collisions, in which the first 'dissipative' encounter $r_p<R_{\rm dissip}$ is also the collision $r_p<R_{\rm col}$,  should follow the time distribution dissipative encounters with a fraction $R_{\rm col}/R_{\rm dissip}$ of these encounters being a collision. Using \eqref{eq:T_col}, the expected time for a clean collision is thus 
\begin{align}\label{eq:T_ccol}
%:eq:T_ccol
&T_{\rm col, est}=\frac{a}{2R_{\rm dissip}}P\sim 1\times 10^7\left(\frac{a}{30\rm AU}\right)^{5/2}\yr\cr
&\times\left(\frac{m_1+m_2}{M_{\odot}}\right)^{-1/2}\left(\frac{R_{\rm dissip}}{4\times 10^9\cm}\right)^{-1}\cr
\end{align}
and is shown to approximately agree with the numerical results in fig \ref{fig:dNdtcol}.

\section{Summary of the numerical integrations}\label{sec:numint}
\subsection{Non-Newtonian corrections}
Two types of calculations are performed. In one, only the Newtonian $1/r^2$ force is included. In the second, GR precession and (non-dissipative) equilibrium tides are included (noted as GR+Tide). GR precession of the WD binary orbit is accounted by including the additional potential energy to the Hamiltonian
\begin{equation}\label{eq:GR_U}
U_{\rm GR}=-3\frac{Gm_1m_2(m_1+m_2)}{c^2r^2}
\end{equation}
which reproduces the long term precession for any eccentricity in the limit of low velocities.
The equilibrium tidal response is included in the quadrupole approximation by the additional potential energy
\begin{equation}\label{eq:Tide_U}
U_{\rm Tide}=-G\frac{k_2m_1^2R_2^5+k_1m_2^2R_1^5}{r^6}
\end{equation}
where $k_1=1,k_2=1$ are the apsidal constants (half the Love numbers), which we conservatively chosen to be unity in all runs.

\subsection{System ensembles}
\label{sec:NumEns} 
The properties of the simulation ensembles used in \sref{sec:NumEns} for the results shown in figures \ref{fig:N_ain},\ref{fig:N_rpout_ain},\ref{fig:dNdRdissip} and the bottom panel of \ref{fig:dNdtcol}, are summarized in table \ref{tab:Simulations}.  The initial and stopping conditions are described in \sref{sec:Numeric}. 

\emph{Inclination selection}
In some runs (denoted 'cut' in table \ref{tab:Simulations}), we avoided the actual calculation of systems with low mutual inclinations $i<85^{\circ}, i>105^{\circ}$ based on a small sample were they were absent. We conservatively assume that all systems that are not numerically integrated, do not lead to a collision.

\emph{Accumulated results for the outer pericenter range $r_{p,\rm out}/a=3-10$}
In figures \ref{fig:N_ain},\ref{fig:dNdRdissip} and the bottom panel of \ref{fig:dNdtcol}, collision fractions are reported for systems with $r_{p,\rm out}$ log-uniformly distributed the range $3-10$. This was calculated by either summing with appropriate waits the fractions achieved at different values of the outer pericenter or (in the case of ensemble F) randomly choosing the outer percenter for each run from a log-uniform distribution.

\emph{Choice of maximal simulation run time $t_{\max}$}
One useful implication of Eq. \eqref{eq:T_ccol} is that there is a natural limit to the amount of orbits $t_{\rm max}/P$ of the numerical integrations that need to be performed to capture the collisions at all semi-major axis values $a$.  There is no need to perform the simulations for times which are much longer than the expected collision (or dissipative events) time given by this equation implying 
\begin{equation}\label{eq:tmax2}
t_{\max}/P\lesssim\frac{a}{2R_{\rm dissip}}\sim 10^5 \frac{a}{30\times AU}.
\end{equation}

On the other hand the number of orbits is limited by the available $5\Gyr$ to   
\begin{equation}\label{eq:tmax1}
t_{\max}/P<3\times 10^7\left(\frac{a}{30\rm AU}\right)^{-3/2}\left(\frac{m_1+m_2}{M_{\odot}}\right)^{1/2}.
\end{equation}
Together, Eqs. \eqref{eq:tmax1} and \eqref{eq:tmax2} imply that runs with $t_{\max}\sim 7\times 10^5 P$ are sufficient to capture the collisions at all separations. We used $t_{\max}\sim 3\times 10^6$. 

\begin{figure}[h]
\includegraphics[scale=0.8]{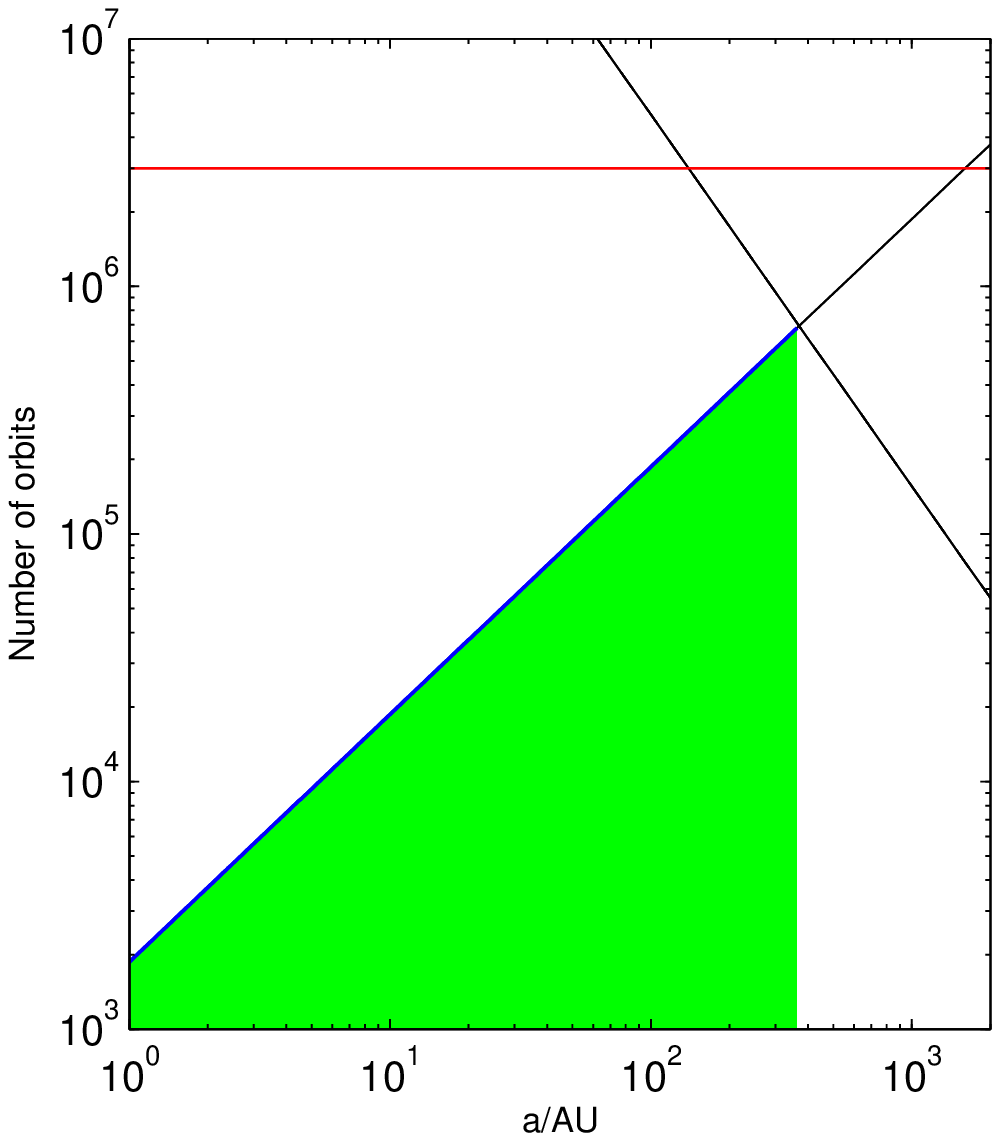}
\caption{Collisions are expected in the green shaded region which is bordered by the expected number of orbits for a collision (blue-black solid line with positive slope Eq. \ref{eq:tmax1}) and the limit $t<5\times 10^9\yr$ (black solid line with negative slope Eq.  \ref{eq:tmax2}).  The amount of orbits integrated in this paper ($t_{\rm max}\sim 2\times 10^6$, red line) is shown to be sufficient to capture collisions at all separations. 
\label{fig:tmax}}
%:fig:tmax
\end{figure}

In practically all simulations (except for 2 simulations with outer apocenter of $r_{p,\rm out}\sim 2,3$ in which the stars are ejected on short times), $t_{\max}$ was chosen to be equal to $2\times 10^6$ orbits. This is justified by the expected collision times shown in figure \ref{fig:tmax} and the collision time distribution from the simulations shown in figure \ref{fig:dNdtcol}.

\begin{table*}
\begin{tabular}{|c|c|c|c|c|c|c|c|c|}
\hline
index&integrator\footnotemark[1]&GR+Tide?\footnotemark[2]&$m_3[M_{\odot}]$&a[AU]&$r_{p,\rm out}/a\footnotemark[3]$&log10($t_{\max}/P$)& \# systems &\#simulations\footnotemark[4]\tabularnewline
 \hline
%$n\footnotemark[1]$ & $\mathcal{L}_{\rm peak}\footnotemark[2]/\mathcal{L}_0$ & $t_{\rm peak}\footnotemark[3]/t_0$  & $\Delta t_{FWHM}/t_0\footnotemark[4]$ & $\mathcal{E}_{\infty}\footnotemark[5]/\mathcal{E}_0=\vt_{\infty}/\vt_0$& $a_t$\footnotemark[5]&$\mathcal{L}_i/\mathcal{L}_0$ \footnotemark[6] & $a_i$ \footnotemark[6] & $b_i$ \footnotemark[6]\tabularnewline
%\hline
A1&PTMT&GR+Tide&0.5&10&2&6.35&500&500\tabularnewline
A2&PTMT&GR+Tide&0.5&10&2.5&6.35&500&500\tabularnewline
A3&PTMT&GR+Tide&0.5&10&3&6.35&500&500\tabularnewline
A4&PTMT&GR+Tide&0.5&10&3.5&6.35&500&500\tabularnewline
A5&PTMT&GR+Tide&0.5&10&4&6.35&500&500\tabularnewline
A6&PTMT&GR+Tide&0.5&10&4.5&6.35&2000&310\tabularnewline
A7&PTMT&GR+Tide&0.5&10&5&6.35&2000&354\tabularnewline
A8&PTMT&GR+Tide&0.5&10&6&6.35&2000&368\tabularnewline
A9&PTMT&GR+Tide&0.5&10&7&6.35&2000&334\tabularnewline
A10&PTMT&GR+Tide&0.5&10&10&6.35&2000&350\tabularnewline
B1&PTMT&-&0.5&-&2&5.35&300&300\tabularnewline
B2&PTMT&-&0.5&-&2.5&6.35&500&500\tabularnewline
B3&PTMT&-&0.5&-&3&5.35&400&400\tabularnewline
B4&PTMT&-&0.5&-&3.5&6.35&500&500\tabularnewline
B5&PTMT&-&0.5&-&4&6.35&500&500\tabularnewline
B6&PTMT&-&0.5&-&4.5&6.35&2000&323\tabularnewline
B7&PTMT&-&0.5&-&5&6.35&1600&253\tabularnewline
B8&PTMT&-&0.5&-&6&6.35&2000&336\tabularnewline
B9&PTMT&-&0.5&-&7&6.35&2000&356\tabularnewline
B10&PTMT&-&0.5&-&8&6.35&2000&363\tabularnewline
B11&PTMT&-&0.5&-&10&6.35&2000&373\tabularnewline
\hline
C1&WH&-&0.5&-&5&6.35&1500&234\tabularnewline
C2&WH&-&0.5&-&6&6.35&1500&276\tabularnewline
C3&WH&-&0.5&-&7&6.35&1500&253\tabularnewline
F&PTMT&GR+Tide&1&100&3-10&6.35&800&800\tabularnewline 
G1&PTMT&-&1&-&2&6.35&300&300\tabularnewline
G2&PTMT&-&1&-&3&6.35&300&300\tabularnewline
G3&PTMT&-&1&-&3.5&6.35&500&500\tabularnewline
G4&PTMT&-&1&-&4&6.35&500&500\tabularnewline
G5&PTMT&-&1&-&4.5&6.35&500&500\tabularnewline
G6&PTMT&-&1&-&5&6.35&500&500\tabularnewline
G7&PTMT&-&1&-&5.5&6.35&2000&345\tabularnewline
G8&PTMT&-&1&-&6&6.35&1600&268\tabularnewline
G9&PTMT&-&1&-&7&6.35&1600&293\tabularnewline
G10&PTMT&-&1&-&8&6.35&1600&284\tabularnewline
G11&PTMT&-&1&-&10&6.35&1600&274\tabularnewline
\hline
\end{tabular}
\footnotetext[1]{PTMT= varying time step,symplectic,second order \cite{Preto99,Mikkola99a,Mikkola99b}, WH=varying time step, non-symplectic with high order (8,6,4) Wisdom Holman \cite{Wisdom91} operator splitting with coefficients taken from \cite{Blanes12}}
\footnotetext[2]{General relativistic and tidal precession using Eqs. \eqref{eq:GR_U},\eqref{eq:Tide_U}}
\footnotetext[3]{In ensembles noted by '3-10', $r_{p,\rm out}/a$ was randomly picked from a log-uniform distribution in the range 3-10} 
\footnotetext[4]{In ensembles denoted 'cut', systems with low inclinations $i<85^{\circ}$ and $i>105^{\circ}$ were not integrated and were conservatively assumed to have no collisions. See text}
\caption{Properties of the Monte Carlo ensembles of systems simulated in \sref{sec:Numeric} and presented in in figures \ref{fig:N_ain},\ref{fig:N_rpout_ain},\ref{fig:dNdtcol_num},\ref{fig:dNdRdissip} and the bottom panel of \ref{fig:dNdtcol}\label{tab:Simulations}}
\end{table*}

\subsection{Integrators}\label{sec:integrators}
The three-body evolution scenarios discussed in this paper occur on very long time scales $\sim 10^6P$, while involving very periastron passages $t_p\sim 10^{-9}(10^6r_p/a)^{3/2}$ implying a range of dynamical times of ~17 orders of magnitude. The following features are helpful for a successful integration of such systems:

\begin{enumerate}
\item Symplectic integration- in symplectic integrators, the change in parameters within each time step is an exact canonical transformation. It turns out that such integrators are stable for very long integrations.
\item Adaptive time step - in order to resolve the short pericenters, an adaptive time step is crucial. 
\item High order and operator splitting - Naturally, convergence is more efficient when high order schemes are used. In the problem considered here, the differential equations can be naturally split into a sum of the large Keplerian terms of the two orbits and the small interaction terms between them (Wisdom-Holman splitting \cite{Wisdom91}). Moreover, each term can be separately analytically integrated. The by giving different waits to different powers of the small parameter at hand (ratio of perturbation to Keplerian terms) very efficient high order schemes can be obtained by alternating the solution of the separate components with universal appropriate time steps.    
\end{enumerate} 

We are not aware of an integration scheme that combines all of these three useful features. We therefore made two competing 'compromises' by using the Preto-Tremain-Mikkola-T (PTMT)\cite{Preto99,Mikkola99a,Mikkola99b} scheme, which is a low (second) order symplectic integrator with an adaptive time step combining 1+2, and a non-symplectic, high order Wisdom-Holman (WH) splitting scheme which combines 2+3. 

We also tried the third possible combination ,namely a symplectic, high order WH integrator combining 1+3 but found it to perform very poorly for the problems at hand. This is somewhat surprising given the fact that at the close pericenter passages, the interaction term is negligible and the Keplerian orbit is advanced analytically, so one would naively imagine that this scheme is very efficient. Unfortunately it isn't, as noted and studied by \cite{Rauch99}. 

\emph{Preto-Tremaine-Mikkola-Tanikawa (PTMT) Integrator}
The PTMT \cite{Preto99,Mikkola99a,Mikkola99b} integrator is the main tool used in this work. The integrator uses the common symplectic leapfrog Kick (change velocities according to forces) and Drift (change positions according to velocities) with a nice trick to allow the use of an adaptive time step without harming the symplectic property.  The time step is chosen as a function $f(U)$ of the total potential energy $U$. It is straight forward to see that the Kick remains a canonical transformation. This is related to the fact that the change in momenta does not depend on the momenta since $U$ is a function of the positions but not the velocities. The Drift is no longer a canonical transformation. The change in positions depends now on the positions through the time step. The Preto-Tremain-Mikkola-Tanikawa trick is to use the conservation of total energy and to replace $U$ with $U\approx E_0-K$ in the expression for the time step when propagating the Kick, where $E_0$ is the initial total energy and $K$ the instantaneous kinetic energy. Now the change in position is a function of momenta only and it is straight forward to check that the transformation is cannonical. The approximation $U\approx E_0-K$ does reduce the accuracy, and we are left with a simple, second order, symplectic integrator with a flexible adaptive time step.

Throughout this work we used a time step  
\begin{equation}\label{eq:dt0}
dt=dt_0\left(\frac{U}{Gm_1m_2/r|_{t=0}}\right)^{-3/2},
\end{equation} 
where $r_{t=0}$ is the initial separation of the WD binary and
\begin{equation}\label{eq:dt00}
dt_0=dt_{00}\left(\frac{a_{t=0}^{3}}{G(m_1+m_2)}\right)^{1/2},
\end{equation}
where $a_{t=0}$ is the initial semi-major axis of the WD binary and $dt_{00}=0.003$. The convergence test shown in the bottom panel of figure \ref{fig:N_ain} used time steps of $dt_{00}=0.01$ (low res, magenta) and $dt_{00}=0.03$ (lower res, cyan) to establish convergence.

\emph{Wisdom Holman (WH) Integrator}
The Wisdom Holman (WH) integrator, uses the natural separation into a Keplerian 'Drift' (D) and interaction 'Kick' (K) terms \cite{Wisdom91} which are separately solved exactly within each step. Appropriate high order convergence is achieved by alternating the integration of each component with appropriate coefficients \cite{Yoshida90}. In this paper we use the following scheme 
\begin{equation}\label{eq:DK}
D_1K_1D_2K_2D_3K_3D_4K_4D_4K_3D_3K_2D_2K_1D_1
\end{equation}
where $D_i$ ($K_i$), implies drifting=propagating the Keplerian orbits (kicking the momenta due to the interaction forces) for a time $dt_i=a_idt$ ($dt_i=b_idt$), where $i=1-4$,  and $a_i,b_i$ were calculated by \cite{Blanes12} and are listed in their table 3. These coefficients result in a method having an error within each step of order $O(\epsilon dt^{8+1}+\epsilon^2 dt^{6+1}+\epsilon^3 dt^{4+1} +\epsilon^4 dt^{2+1})$ where $\epsilon$ is the small parameter quantifying the strength of the interaction perturbation as compared to the Keplerian term. This generalized order is denoted $(8,6,4)$. 

Given that $K$ and $D$ are canonical transformations (any Hamiltonian time propagation is) the advancement of one step is canonical and such methods are usually used as high order symplectic integrators. Unfortunately, as explained above, our integration is not symplectic due to our use of an adaptive time step to allow the pericenter passages to be resolved. Indeed, the dependence of the time step on the coordinates makes the transformation non cannonical. We use the following space dependent time step,
\begin{equation}\label{eq:dt0WH}
dt=dt_{0,WH}\left(\frac{r}{a_{t=0}}\right)^{3/2},
\end{equation}
which is updated after each completion of the Drift-Kick sequence \eqref{eq:DK}. In the integrations presented here, a time step coefficient $dt_{0,WH}=0.1$ is used. For this choice of time resolution, the positions and times of all pericenters in the example shown in figure \ref{fig:example_r_t_a}, are converged to an accuracy better than $10^{-2}$, as confirmed by runs with higher time resolutions.

%In the common leapfrog integrator, the Drift,
%\begin{align}
%&x_i'=x_i+dt*p_i/m\cr 
% &p_i'=p_i\cr
%\end{align}
% and Kick, 
%\begin{align}
%&p_i'=p_i+dt*F_i(x)\cr 
% &x_i'=x_i\cr
%\end{align}
%are symplectic since any transformation of the form 
%
%\begin{align}\label{eq:SimplecticT}
%&x_i'=x_i+f_i(p)\cr 
% &p_i'=p_i\cr
% \end{align}
% is canonical if $f$ satisfies $\pr_{p_i}f_j=\pr_{p_j}f_i$ (the roles of x,and p change in the drift and kick). This is easy to see by directly verifying that the symplectic tensor $\eta_{x_i,p_j}=-\eta_{p_i,x_j}=\delta_{i,j},\eta_{x_i,x_j}=\eta_{p_i,p_j}=0$  transforms to itself.
%
%If we choose a space dependent time step $g(x^2)$  
% 
%  
%are both of the form \ref{eq:SimplecticT}


\begin{thebibliography}{01}
\bibitem[Howell(2011)]{Howell11} Howell, D.~A.\ 2011, Nature 
Communications, 2,  350

\bibitem[Rosswog et al.(2009)]{Rosswog09} Rosswog, S., Kasen, D., 
Guillochon, J., \& Ramirez-Ruiz, E.\ 2009, \apjl, 705, L128  %WDWD Numeric 
\bibitem[Raskin et al.(2009)]{Raskin09} Raskin, C., Timmes, 
F.~X., Scannapieco, E., Diehl, S., \& Fryer, C.\ 2009, \mnras, 399, L156 %WDWD Numeric 
\bibitem[Raskin et al.(2010)]{Raskin10} Raskin, C., Scannapieco, 
E., Rockefeller, G., et al.\ 2010, \apj, 724, 111 %WDWD Numeric 
\bibitem[Hawley et al.(2012)]{Hawley12} Hawley, W.~P., 
Athanassiadou, T., \& Timmes, F.~X.\ 2012, \apj, 759, 39 %WDWD Numeric 
\bibitem[Lor{\'e}n-Aguilar et al.(2010)]{Loren10} 
Lor{\'e}n-Aguilar, P., Isern, J., 
\& Garc{\'{\i}}a-Berro, E.\ 2010, \mnras, 406, 2749  %WD-WD numeric fail

\bibitem[Cappellaro et 
al.(1999)]{Cappellaro99} Cappellaro, E., Evans, R., \& Turatto, M.\ 1999, \aap, 351, 459 

\bibitem[Horiuchi 
\& Beacom(2010)]{Horiuchi10} Horiuchi, S., \& Beacom, J.~F.\ 2010, \apj, 723, 329 %SNe Ia rate

\bibitem[Lidov(1962)]{Lidov62} Lidov, M.~L.\ 1962, Planetary and Space Science, 9, 719 
\bibitem[Kozai(1962)]{Kozai62} Kozai, Y.\ 1962, Astron. J., 67, 591 
%\bibitem[Heisler \& Tremaine(1986)]{Heisler86} Heisler, J., \& Tremaine, S.\ 1986, Icarus, 65, 13 
\bibitem[Kiseleva et al.(1998)]{eggleton} Kiseleva, L.~G., 
Eggleton, P.~P., \& Mikkola, S.\ 1998, \mnras, 300, 292
\bibitem[Blaes et al.(2002)]{blaes} Blaes, O., Lee, M.~H., 
\& Socrates, A.\ 2002, \apj, 578, 775 
\bibitem[Wu 
\& Murray(2003)]{wumurray} Wu, Y., \& Murray, N.\ 2003, \apj, 589, 605
\bibitem[Fabrycky 
\& Tremaine(2007)]{fabrycky} Fabrycky, D., \& Tremaine, S.\ 2007, \apj, 669, 1298 
\bibitem[Dong et al.(2012)]{dong} Dong, S., Katz, B., 
\& Socrates, A.\ 2012, arXiv:1201.4399 
\bibitem[Socrates et al.(2012)]{socrates} Socrates, A., Katz, 
B., Dong, S., \& Tremaine, S.\ 2012, \apj, 750, 106
\bibitem[Holman et al.(1997)]{Holman97} Holman, M., Touma, J., 
\& Tremaine, S.\ 1997, \nat, 386, 254 

\bibitem[Ford et al.(2000)]{Ford00} Ford, E.~B., Kozinsky, B., 
\& Rasio, F.~A.\ 2000, \apj, 535, 385

\bibitem[Touma et al.(2009)]{Touma09} Touma, J.~R., Tremaine, 
S., \& Kazandjian, M.~V.\ 2009, \mnras, 394, 1085 

\bibitem[Naoz et al.(2011a)]{Naoz11} Naoz, S., Farr, W.~M., 
Lithwick, Y., Rasio, F.~A., \& Teyssandier, J.\ 2011, \nat, 473, 187 

\bibitem[Katz et al.(2011)]{Katz11} Katz, B., Dong, S., 
\& Malhotra, R.\ 2011, Physical Review Letters, 107, 181101 

\bibitem[Lithwick 
\& Naoz(2011)]{Lithwick11} Lithwick, Y., \& Naoz, S.\ 2011, \apj, 742, 94 

\bibitem[Bode \& Wegg (2012?)]{Bode12} Bode N., \& Wegg C. \ 2012, in preparation , more information in http://www.chriswegg.com

\bibitem[Thompson(2011)]{Thompson11} Thompson, T.~A.\ 2011, \apj, 
741, 82 
\bibitem[Shappee 
\& Thompson(2012)]{Shappee12} Shappee, B.~J., \& Thompson, T.~A.\ 2012, arXiv:1204.1053 

\bibitem[Katz \& Dong (2012)]{Katz12} Katz, B. \& Dong, S. \ 2012, in preparation.

\bibitem[Wisdom 
\& Holman(1991)]{Wisdom91} Wisdom, J., \& Holman, M.\ 1991, \aj, 102, 1528 
\bibitem[Preto 
\& Tremaine(1999)]{Preto99} Preto, M., \& Tremaine, S.\ 1999, \aj, 118, 2532 
\bibitem[Mikkola 
\& Tanikawa(1999)]{Mikkola99a} Mikkola, S., \& Tanikawa, K.\ 1999, \mnras, 310, 745 
\bibitem[Mikkola 
\& Tanikawa(1999)]{Mikkola99b} Mikkola, S., \& Tanikawa, K.\ 1999, Celestial Mechanics and Dynamical Astronomy, 74, 287 
\bibitem[Blanes et al.(2012)]{Blanes12} Blanes, S., Casas, F., 
Farres, A., et al.\ 2012, arXiv:1208.0689 

\bibitem[Liebert et al.(2005)]{Liebert05} Liebert, J., Bergeron, 
P., \& Holberg, J.~B.\ 2005, \apjs, 156, 47 %WD formation rate is 1e-12 /yr/pc^3
\bibitem[Lidov 
\& Ziglin(1976)]{Lidov76} Lidov, M.~L., \& Ziglin, S.~L.\ 1976, Celestial Mechanics, 13, 471 
\bibitem[Peters(1964)]{Peters64} Peters, P.~C.\ 1964, Physical 
Review, 136, 1224 
\bibitem[Naoz et al.(2011b)]{Naoz11b} Naoz, S., Farr, W.~M., 
Lithwick, Y., Rasio, F.~A., \& Teyssandier, J.\ 2011, arXiv:1107.2414 

\bibitem[Shara 
\& Regev(1986)]{Shara86} Shara, M.~M., \& Regev, O.\ 1986, \apj, 306, 543 

\bibitem[Soker 
\& Tylenda(2006)]{Soker06} Soker, N., \& Tylenda, R.\ 2006, \mnras, 373, 733 
\bibitem[Yoshida(1990)]{Yoshida90} Yoshida, H.\ 1990, Physics 
Letters A, 150, 262 

\bibitem[Piro(2011)]{Piro11} Piro, A.~L.\ 2011, \apjl, 740, 
L53 %% observational Q in WD

*Bahcal Fellow, Einstein Fellow
\end{thebibliography}
\end{document}